\documentclass[
 a4paper,twoside,10pt,superscriptaddress,aps,
twocolumn,
 tightenlines,
 amsmath,amssymb,footinbib,
]{revtex4}

\usepackage{graphicx}
\usepackage{float}
\usepackage[usenames, dvipsnames]{color}
\usepackage{bm,soul}
\usepackage[colorlinks=true,linkcolor=blue,citecolor=blue,urlcolor=blue,breaklinks=true]{hyperref}
\newcommand{\ket}[1]{\left| #1 \right>} 

\begin{document}

\title{Interfering trajectories in experimental quantum-enhanced stochastic simulation}

\author{Farzad \surname{Ghafari}}
\email{farzad.ghafari@griffithuni.edu.au}
\affiliation{Centre for Quantum Dynamics, Griffith University, Brisbane, 4111, Australia}

\author{Nora \surname{Tischler}}
\affiliation{Centre for Quantum Dynamics, Griffith University, Brisbane, 4111, Australia}

\author{Carlo Di Franco}
\affiliation{School of Physical and Mathematical Sciences, Nanyang Technological University, 639673, Singapore}
\affiliation{Complexity Institute, Nanyang Technological University, 639673, Singapore}

\author{Jayne Thompson}%
\affiliation{Centre for Quantum Technologies, National University of Singapore, 117543, Singapore}

\author{Mile Gu}%
\email{mgu@quantumcomplexity.org}
\affiliation{School of Physical and Mathematical Sciences, Nanyang Technological University, 639673, Singapore}
\affiliation{Complexity Institute, Nanyang Technological University, 639673, Singapore}
\affiliation{Centre for Quantum Technologies, National University of Singapore, 117543, Singapore}

\author{Geoff J. Pryde}%
\email{g.pryde@griffith.edu.au}
\affiliation{Centre for Quantum Dynamics, Griffith University, Brisbane, 4111, Australia}

\begin{abstract}
Simulations of stochastic processes play an important role in the quantitative sciences, enabling the characterisation of complex systems. Recent work has established a quantum advantage in stochastic simulation, leading to quantum devices that execute a simulation using less memory than possible by classical means. To realise this advantage it is essential that the memory register remains coherent, and coherently interacts with the processor, allowing the simulator to operate over many time steps. Here we report a multi-time-step experimental simulation of a stochastic process using less memory than the classical limit. A key feature of the photonic quantum information processor is that it creates a quantum superposition of all possible future trajectories that the system can evolve into. This superposition allows us to introduce, and demonstrate, the idea of comparing statistical futures of two classical processes via quantum interference. We demonstrate interference of two 16-dimensional quantum states, representing statistical futures of our process, with a visibility of $0.96 \pm 0.02$.
\end{abstract}
\maketitle
\section*{Introduction}
Many of the most interesting phenomena are complex---whether in urban design, meteorology or financial prediction, the systems involved feature a vast array of interacting components. Predicting and simulating such systems often requires the use of a prohibitive amount of data, evincing a pressing need for more efficient tools in algorithmic modelling and simulation.

Quantum technologies have shown the potential to dramatically reduce the amount of working memory required to simulate stochastic processes~\cite{Gu2012,Mahoney2016}. By tracking information about past observations directly within quantum states, a quantum device can replicate the system's conditional future behaviour, using less memory than the provably optimal classical limits. The key to achieving a quantum memory advantage is maintaining coherence of the quantum memory during the simulation process, enabling the encoding of relevant past information into non-orthogonal quantum states. This memory reduction comprises a new application of quantum processing, complementary to
computational speedup \cite{Lloyd1996}, cryptography \cite{Bennett1984}, sensing~\cite{Giovannetti2004,Slussarenko2017} and phase estimation~\cite{Xiang2011}.


This advantage was first illustrated for simulating a particular stochastic process, where past information was encoded within non-orthogonal polarisation states of a single photon~\cite{Palsson2016}. The scheme, however, maintained quantum coherence over only a single simulation cycle. This limitation meant that the resulting simulator exhibited a memory advantage only when simulating a single time step. To simulate multiple time steps, such a device required relevant information to be transferred to classical memory between time steps, negating any quantum advantage. 

Here we develop a quantum simulator that overcomes this limitation, such that it exhibits a memory advantage when simulating multiple time steps.  As an important additional benefit, our device enables us to create a quantum superposition over all potential future outcomes of a process. We illustrate that such an output lets us estimate the distinguishability in the statistical futures of two stochastic systems via quantum interference. Our experimental approach makes use of temporal (time-bin) encoding in an optical system to experimentally realise a quantum simulation over three consecutive steps, generating a coherent superposition over the process's potential future trajectories. We then implement two such quantum simulations in parallel, simultaneously generating superpositions over the trajectories for each of two independent systems. Experimentally, this corresponds to using our quantum simulators to produce and control high-dimensional quantum states. These are interfered, allowing estimation of how well the corresponding statistical futures coincide.

\section*{Results}

\subsection*{Framework and tools}

\begin{figure}
	\centering
	\includegraphics[width=1\linewidth]{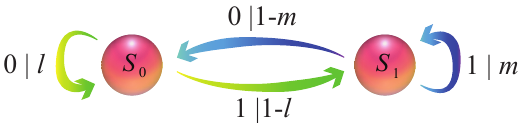}
	\caption{\textbf{Perturbed coin.} The process that we study here is a coin with two outcomes, 0 and 1. The transition probabilities, $T_{ij}$, between different outcomes are determined by $ l $ and $ m $ for $ i, j \in\{0,1\} $.  The optimal classical model uses the causal states ($ \{ S_i\} $) depicted in the circles. There is a simple mapping from the past of the process to the relevant causal state: the last outcome from the coin determines the input causal state. Arrows, with the associated expressions $j|\ T_{ij}$, represent the transitions from causal states $ S_i $ to $ S_j $ with probability $T_{ij}$, emitting the classical outcome  $j$. In the quantum model, the causal states become quantum states, $\{|S_i\rangle\}$.}
	\label{fig:process}
\end{figure}

In this work, we study a simple stochastic process known as the perturbed coin~\cite{Gu2012}. It consists of a binary random variable that represents the state of a coin (0 corresponds to heads, and 1 to tails) inside a box. At each time step, the box is perturbed, causing the coin to flip with some probability. Afterwards, the state of the coin is emitted. In general the coin may be biased, so the probability of remaining in heads, $l$, can differ from the probability of remaining in tails, $m$, as presented in Fig.~\ref{fig:process}. Repetition of this procedure generates a string of 0s and 1s, whose statistics define the perturbed coin process.

Any device that seeks to replicate correct future statistics must retain relevant past information in a memory. This involves a prescription for configuring its memory in an appropriate state for each possible observed past, such that systematic actions on this memory recover a sequence of future outputs that are faithful to conditional future statistics. The amount of past information stored in memory is quantified by the Shannon entropy $C = - \sum_s d_s \log d_s$, where $d_s$ is the probability that the memory is in state $s$ and the logarithm is in base 2. The minimal possible memory required, $C_{\mathrm{\mu}}$, is known as the statistical complexity, and is an important measure of structure in complexity science~\cite{Grassberger1986,Crutchfield1989,Shalizi2001,Crutchfield2009}. For the perturbed coin (Fig.~\ref{fig:process}), the minimal information required about the past is the current state of the coin. This induces a statistical complexity of $C_{\mu} = -q \log q - (1 - q) \log (1-q)$, where $q$ represents the probability that the last outcome was heads (see Eq. (\ref{statisticalcomplexity}) in Methods).

\begin{figure}
	\centering
	\includegraphics[width=1\linewidth]{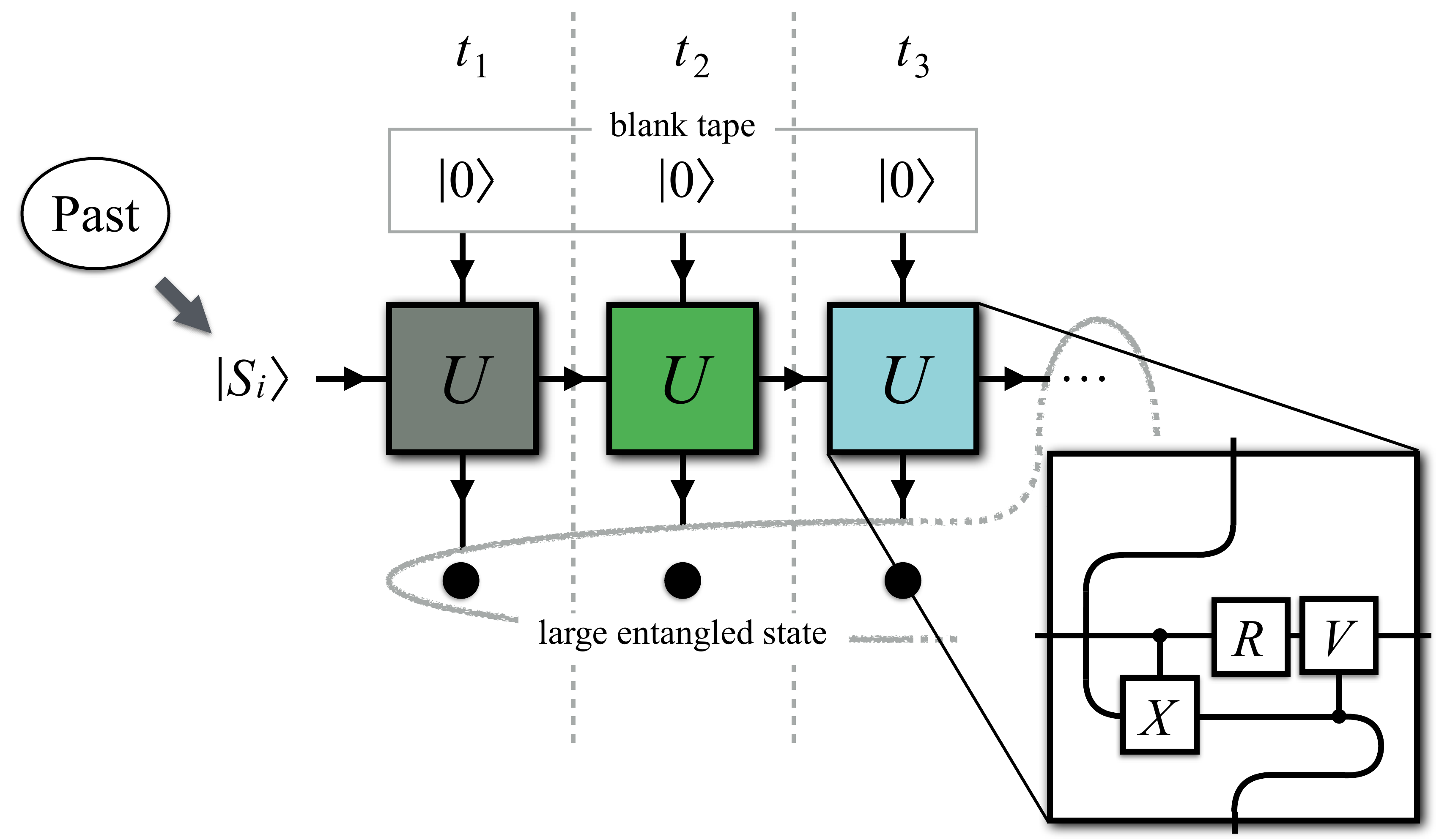}
	\caption{\textbf{Conceptual representation of the multiple-step quantum simulation of a perturbed coin.} The memory system of the simulator is initialised in  a qubit state  $\ket{S_i}$, where $i\in\{0,1\}$ depends on the past of the process. The ancillary qubits are all initialised in a fixed known state $\ket{0}$--- the logical zero state---and thus contain no information about the past of the process. At each time step $t_k$, the simulator interacts with the $k$th ancilla through the same unitary operator $ U $. The inset shows how we implement the relevant unitary operator. The gates are a controlled-$ X $, a single-qubit rotation R such that $R\ket{0}=\ket{S_0}$, and a controlled-$ V $ such that $VR\ket{1}=\ket{S_1}$. This sequence of interactions results in an entangled internal state that includes all the ancillary qubits and the memory state of the simulator. Measuring the ancillas samples the statistical distribution of the process, and at the same time the internal state of the simulator collapses into the correct memory state required for further simulation steps.}
	\label{fig:diagram}
\end{figure}

A quantum simulator can further reduce memory requirements by encoding the two possible outcomes of the process into mutually non-orthogonal states. Future statistics are then generated by a series of unitary interactions, ensuring that this entropic advantage is maintained at all times during simulation~\cite{Binder2018}. For the case of the perturbed coin, the quantum simulator can be implemented as shown in Fig.~\ref{fig:diagram}. The state of the machine encodes relevant information about past outcomes---here, the state of the coin after the last step. It is represented as one of two states, $\ket{S_0}$ or $\ket{S_1}$, of a quantum system that sequentially interacts with ancillary systems. Each interaction corresponds to a time step of the stochastic process. All the ancillary systems start in a fixed state, and therefore do not contain any information. The sequence of interactions produces an entangled state. Measuring the ancillary systems after the desired number of steps provides a sample of the statistics.

\subsection*{Experimental implementation}

Motivated by recent realisations of quantum walks in linear optical setups with time-bin encoding \cite{Schreiber2010,Schreiber2012,Jeong2013,Boutari2016}, we implement the memory system and multiple ancillas---here, corresponding to three time steps---by encoding on a single photon. The ancillas, which can be read to obtain the classical outcomes of the process, are encoded in the arrival time of the photon, and the memory state of the simulator is encoded in its polarisation. Thus, for a simulation of $M$ time steps, a $2^M$-dimensional system corresponding to $2^M$ different photon arrival times replaces $M$ distinct ancillary photons. Instead of measuring the classical outcome at each time step, our quantum information processor keeps the photon and builds up a superposition in a high-dimensional Hilbert space; in our case $M=3$, and the output of the simulator is 16 dimensional (8 arrival time modes $\times$ 2 polarisation modes). The associated memory cost during this process does not increase since all operations remain unitary---and thus conserve entropy. Of course, distinct ancilla qubits could be used instead, but encoding in multiple degrees of freedom provides a convenient, effective and high-fidelity approach for small- to medium-sized photonic systems.

Our experiment demonstrates that high-dimensional (here 16-dimensional) quantum states can be encoded and manipulated in photonic temporal and polarisation modes with high fidelity~\cite{Franson1989,Kwiat1993}. This complements other related works involving hybrid optical states using spatial (path and optical orbital angular momentum) and polarisation modes~\cite{takeuchi2000experimental,ma2009experimental,nagali2010generation,Zhang2016}. It also substantiates the oft-repeated claim that combining different photonic encodings~\cite{Kwiat1997,Barreiro2005} is a practical tool for various quantum information tasks, for example studying the remote preparation of entangled states~\cite{barreiro2010remote}, complementarity~\cite{nogueira2010interference}, Bell inequalities~\cite{valles2014generation,ma2009experimental,dada2011experimental}, quantum key distribution implementations~\cite{takemoto2015quantum} and complete optical Bell state analysers~\cite{walborn2003hyperentanglement,wei2007hyperentangled}.

\begin{figure}[h!]
	\centering
	\includegraphics[width=1\linewidth]{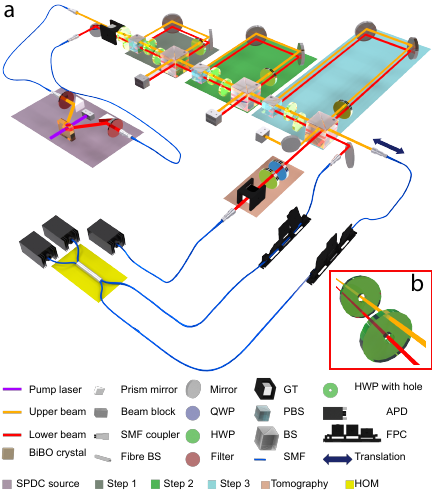}
	\caption{\textbf{Experimental setup.} \textbf{a} Single photons are generated from a degenerate spontaneous parametric down-conversion (SPDC) source pumped by a 410 nm continuous-wave laser. After filtering the generated photons with $(820 \pm 1.5)$ nm bandpass filters, photons in the lower beam (red) and upper beam (orange) are separately prepared in their respective input states, $\ket{S_0}$ or $\ket{S_1}$, using half-wave plates (HWP). Polarisation qubits, one from each beam, are used as memory states for the simulation of two separate and potentially different processes $ \Pi_1 $ (red) and $ \Pi_2 $ (orange). (For the first experiment described in the text, where only one process is required, the second SPDC output (orange) is sent straight to a heralding detector, rather than through the apparatus.) To implement the three-step simulation, three processor blocks are built---labelled Step 1, Step 2, and Step 3. In each step, path and arrival time modes are also employed to realise the relevant physical operation, as explained further in Methods. The output of one of the simulators (lower beam) is used to perform the polarisation tomography and to measure the arrival times of each photon in order to sample the statistical future. To measure the overlap of the future statistics of two processes, both photons are used, and the other outputs of the third beam splitter (BS) are interfered in a fibre BS (yellow box). An automated translation stage is used to move one of the couplers in order to vary the relative delay between the single-photon wave packets. Avalanche photodiodes (APD) and a single-photon counting module are used to count the photons. SMF stands for single-mode fibre, QWP for quarter-wave plate, GT for Glan-Taylor prism, PBS for polarising beam splitter, and FPC for fibre polarisation controller. \textbf{b} The inset shows a close-up of two vertically-separated beams passing through two HWPs with holes, each of which only acts on one of the beams.}
	\label{fig:setup}
\end{figure}

Our first task consists of performing the quantum simulation of the perturbed coin. In particular, we seek to verify that the simulator samples from the correct statistical distributions, and to demonstrate the memory advantage due to quantum encoding. The experimental setup is shown in Fig.~\ref{fig:setup}. We generate degenerate pairs of single photons through spontaneous parametric down-conversion. One of the photons (depicted as the red, lower beam in the figure) is prepared in the state $\ket{S_0}$ or $\ket{S_1}$, depending on the past of the process. It then passes through three sequential blocks, which represent the three time steps being simulated. In each block, the short and long paths correspond to outcomes $ 0 $ and $ 1 $, respectively (details in Methods).  For the simulation, only one of the photons (the red beam) is used, and the other photon (orange beam in the figure) is not used except as a herald, and is measured immediately after generation (for this task, it does not go through the apparatus as shown in the figure). We then estimate the polarisation state of the red-beam photon in the tomographic reconstruction at the end of the third block, and also measure its arrival time (using the orange-beam photon as a reference). In this way, we obtain the probability distribution of the stochastic process as simulated by our quantum information processor, together with the final memory state of our simulator, which is needed for further simulation steps.

\subsection*{Experimental results}

\begin{figure}
	\centering
	\includegraphics[width=1\linewidth]{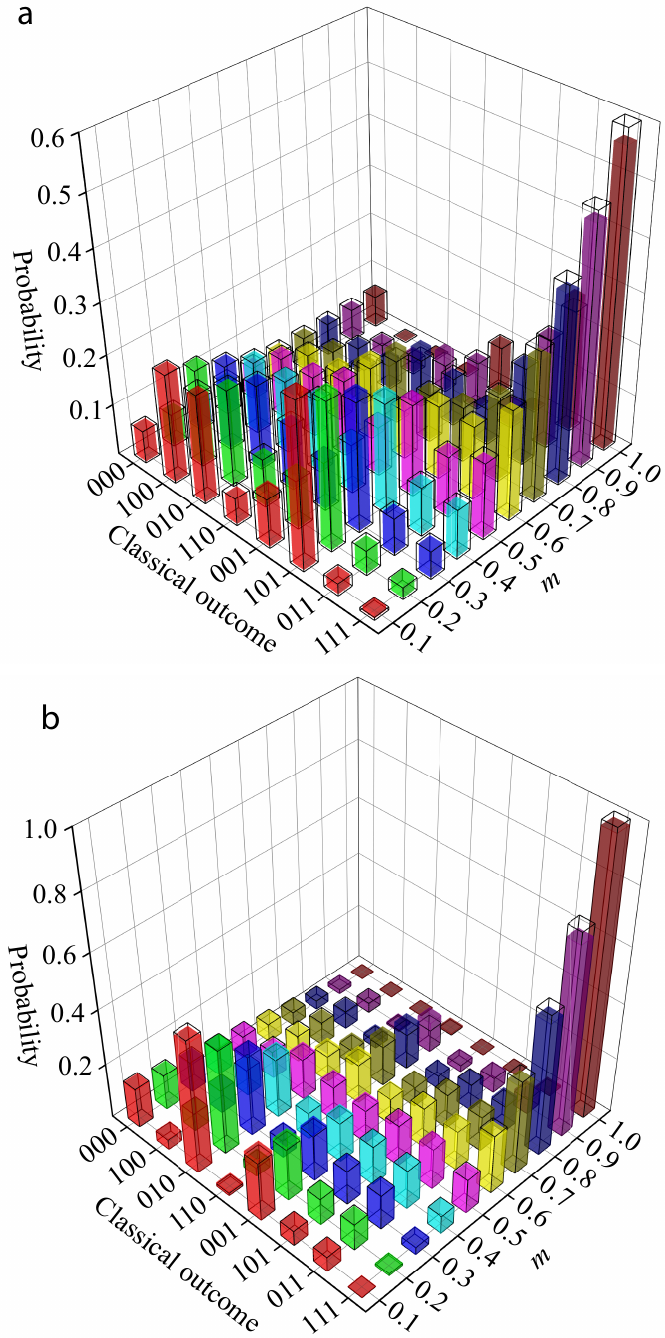}
	\caption{\textbf{Sampling of statistical futures.} \textbf{a} The coloured bars show the measured probability of different outcomes for the three simulation steps, when the initial state is $ |S_0\rangle $ with $ l=0.4 $, and for discrete values of $ m $ in the range $ 0.1 $ to $ 1.0 $. The transparent bars show the theoretically-calculated probabilities for the corresponding process. \textbf{b} The sampled future of the same process, when the initial state is $|S_1\rangle $.  Uncertainties, due to the Poissonian distribution of photon counts, are so small that they are not visible in the graphs---therefore, they are not depicted. Note that the classical probability distribution is determined by the process parameters $l$ and $m$, as well as the initial causal state. For example, if the last outcome of the coin is 1, the quantum simulator is initialised in state $|S_1\rangle$. The conditional probability of subsequently observing $111$ is then $m^3$. For a fixed $l =0.4$  and increasing $m$, the average probability of getting 1 in the simulation thus rises accordingly. This can be seen by the higher columns in the right corners of both graphs.}
	\label{fig:Classical Probs}
\end{figure}

The experimentally determined outcome probabilities are shown in Fig.~\ref{fig:Classical Probs}, and are close to the expected theoretical values. The main discrepancies with theory are due to small differences between nominally identical polarisation elements, and the non-identical single-mode-fibre coupling efficiency of photons taking different paths through the simulator. In order to evaluate how well they agree, we calculate the (classical) fidelity~\cite{Book-Nielsen2010} for each set of parameters and initial conditions that we have simulated in our experiment. All the values obtained for this (classical) fidelity are larger than $0.991$. 
Typical uncertainties are around $0.001$.

\begin{figure*}
	\centering
	\includegraphics[width=1\linewidth]{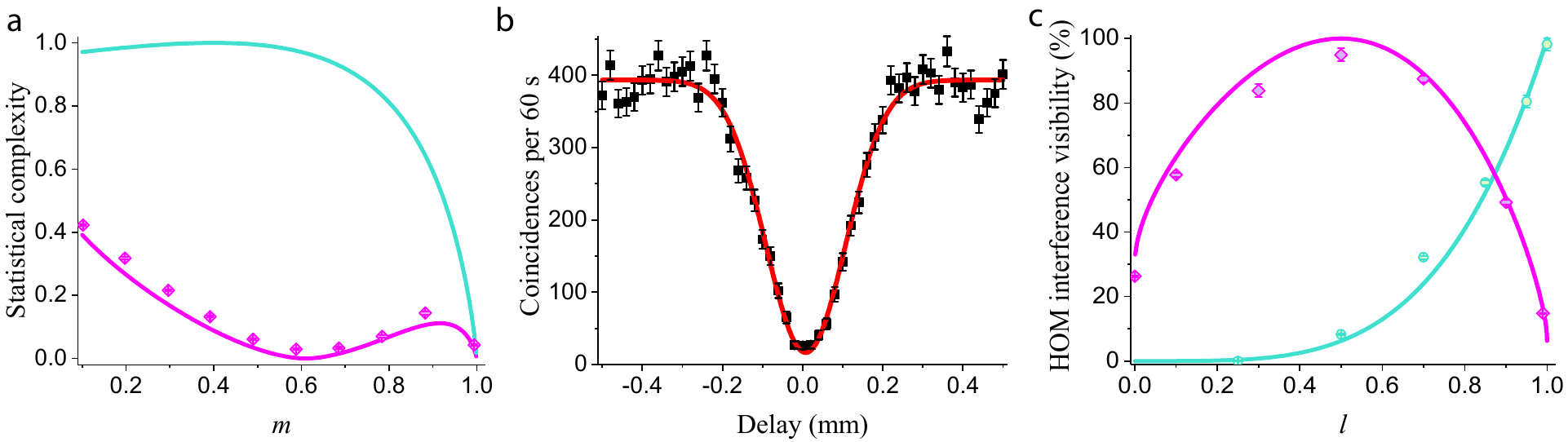}
	\caption{\textbf{Experimentally determined complexity and interference visibility.} \textbf{a} The quantum statistical complexity $C_q$ of our quantum simulator, as the probability of remaining in tails during perturbation, $m$, is varied. The probability of remaining in heads, $ l $, is fixed at $ 0.4 $. Due to small experimental imperfections the actual implemented values of $ l $ and $ m $ deviate slightly from the nominal round values (see Methods for more details). Data points are experimental measurements of $ C_q $, and the magenta and turquoise curves are theoretical estimations for the quantum and classical complexities $C_q$ and $C_{\mu}$, respectively. \textbf{b} Two-photon interference of the superpositions of future trajectories in two implemented stochastic processes, $ \Pi_1 $ and $ \Pi_2 $, with $ \Pi_1=\Pi_2 $ such that $ l=0.5 $ and $ m=0.5 $ and thus $ |S_0\rangle = |S_1\rangle$. Since $ \Pi_1=\Pi_2 $, an interference visibility of $ 100\% $ is theoretically expected, while fitting the experimental data yields a visibility of $ 0.96 \pm 0.02$. In the graph, the number of measured twofold coincidences is depicted versus the relative delay between single-photon wave packets. \textbf{c} Magenta and turquoise elements (points---experiment; curves---theory) show the comparison of the statistical futures from two stochastic processes by two-photon interference visibility. In each case, one process ($\Pi_1$) is fixed, and the other process ($\Pi_2$) has fixed $m$ but varying $l$. Magenta represents interference of the output states of the simulators for $ \Pi_1 $ ($ |S_0\rangle $ is the input memory state, $ l=0.5$, and $m=0.5 $) with $ \Pi_2 $ ($ |S_0\rangle $ is the input memory state, $ m=0.5 $, and $ l $ varying). Turquoise represents interference of the output states of the simulators for $ \Pi_1 $ ($ |S_0\rangle $ is the input memory state, $ l=1.0$, and $m=1.0 $) with $ \Pi_2 $ ($ |S_0\rangle $ is the input memory state, $ m=0.5 $, and $ l $ varying). Uncertainties, from the Poissonian distribution of photon counts, are so small that they are barely visible in some of the graphs.}
	\label{fig:result}
\end{figure*}

To compare the use of quantum and classical resources, we use $C_q$, the quantum counterpart to the classical statistical complexity (the entropy of the memory register of the quantum simulator), which quantifies the memory requirement of the quantum simulator. We thus calculate $C_q$ for this process (details in Methods).
The experimental results are shown in Fig.~\ref{fig:result}a. The corresponding classical statistical complexity is also shown for the sake of comparison, demonstrating that quantum resources dramatically reduce the amount of memory needed for simulating a multi-step stochastic process.

To guarantee that the quantum  memory advantage is maintained at all stages of the simulation process, we require the internal dynamics to be close to (ideally, completely) unitary. We can verify this by demonstrating the coherence of the output state that includes all the ancillary qubits and the memory state of the simulator.
We observe this coherence via two-photon quantum interference. We use the complete setup of Fig.~\ref{fig:setup}, where the photon depicted by the orange path is no longer measured after generation (as done previously), but also goes through the apparatus. Both the photons pass independently through the three sequential blocks, with each experiencing nominally the same optical elements (although different settings are possible). If the coherence between the different time bins and polarisations exploited in our simulation is maintained, we expect a complete interference, which means that the visibility ideally should be unity. The result in Fig.~\ref{fig:result}b shows a visibility of $0.96 \pm 0.02$ for the case where the theoretical output states of the apparatus are uniform superpositions of all time bins and polarisations (which is the scenario where the highest discrepancy from the ideal visibility would be expected as it is most susceptible to imperfections). The high value obtained here indicates that our simulator is (almost) implementing a unitary operator, and the entropy of our system does not significantly increase throughout the simulation process. This requirement is essential for preserving the quantum memory advantage. Moreover, apart from the specific application of this apparatus to simulate classical stochastic processes, this result is also significant in a more general context, since it demonstrates the interference of two discrete high-dimensional states with an extremely high visibility~\cite{Zhang2016}.

Modifying this experimental setup allows us to compare two different processes, $\Pi_1$ and $\Pi_2$. Clearly, one way to perform such a statistical comparison is to consider each process individually, and sample its outcomes to reconstruct the corresponding distribution. These two reconstructed distributions can then be compared. However, we notice that in our quantum simulation, all the information about the future statistics is already encoded in the state that exists in our apparatus. Thus, we do not need to collapse the superposition of possible outcomes by sampling, instead we can exploit this superposition for our task of comparing the future of processes. In particular, by simultaneously running quantum simulations of processes $\Pi_1$ and $\Pi_2$ in parallel and interfering the resulting output states, we can estimate the overlap of their future statistics.

In our experiment, we realise different processes by applying different operations to the two photons (red beam and orange beam) in the three blocks of the setup in Fig.~\ref{fig:setup}. To implement the parameters of each process separately, we use half-wave plates with holes, which allow us to change the polarisation of one beam without affecting the other. We fix one of the processes and change the other process gradually. As the parameters defining the processes become increasingly similar, the two output probability distributions overlap more. This is reflected in the experiment by a higher visibility value, showing how the comparison between two sets of future statistics can be evaluated via interference visibility. Results are shown in Fig.~\ref{fig:result}c, where the experimental values are close to theoretical predictions. However, there remain slight discrepancies because of experimental imperfections such as small spatial and polarisation mode mismatches. These techniques could be adapted to attain a quantum advantage in estimating the distance between two normalised vectors~\cite{Kumar2017}, which plays an essential role in machine learning tasks such as image recognition~\cite{Book-Shalev-Shwartz2014}.

\section*{Discussion}

Our multi-step photonic implementation of a stochastic simulation has verified the memory advantage available with quantum resources. We have demonstrated that it is possible to maintain this advantage at all stages of the simulation by preserving quantum coherence, as opposed to previous experiments~\cite{Palsson2016,Jouneghani2017,Ghafari2018}. Further, we have shown that superpositions of process outcomes can be interfered. These techniques have the potential to reduce memory requirements in simulations of stochastic processes and to provide tools for advances in quantum machine learning and communication complexity.

The time-bin-encoding techniques in our experiment can be extended to other small- and medium-scale simulations by expanding the number of time bins. For example, $ 10^8 $ time-bin modes have been realized in the context of communication complexity~\cite{Xu2015}. However, the number of bins does not scale efficiently with the number of qubits, and thus very-large-scale simulations are not possible with this encoding. This is not a fundamental problem, as the concepts that we demonstrate can be equivalently implemented in other photonic encodings or in other qubit systems. Our current demonstration also uses non-deterministic (post-selected) mode recombination at certain beam splitters within the circuit. This implementation is convenient, but not necessary and thus not a fundamental limitation: a deterministic multi-step simulator could be realised with a step-dependent delay mechanism--- for instance, a controlled fast switch connected to fibre paths of different lengths.

The comparison of future statistics has direct relation to other protocols, such as quantum fingerprinting and state comparison in communication complexity~\cite{Xu2015,Kumar2017}. Fingerprinting involves estimating the distance between two vectors, where the resource to be minimised is the amount of communication. For the comparison of two vectors, quantum mechanics can reduce the amount of communication required beyond classical limits. In the quantum protocol, Alice and Bob perform a SWAP test ---a quantum information primitive which compares two arbitrary states. Two-photon interference is known to be equivalent to a SWAP test \cite{Garcia-Escartin2013}. Our comparison of futures can be cast as a similar problem. In this case, the task would be for Alice and Bob, who each have their future statistics from potentially different processes, to compare the two statistical futures \cite{Kumar2017}. In principle, for very high-dimensional Hilbert spaces,  a comparison of statistical futures via two-photon interference can achieve a quantum advantage in communication complexity. The comparison of two vectors is also an important component of many machine learning tasks, and thus a similar advantage could extend to more general settings like speech recognition~\cite{Book-Shalev-Shwartz2014}.

\section*{Methods}
\subsection*{Theoretical background}

A discrete-time stochastic process is generally described by a joint probability distribution, $ p(\overleftarrow{X},\overrightarrow{X}) $, where $\overleftarrow{X}= ...,\, X_{-1},\,X_0 $ ($ \overrightarrow{X}= X_1,\,X_2, \,... $) denotes the random variables that govern the statistics of past (future) observations. Each past (future) configuration of the random process is denoted by ${\overleftarrow{x}} $ (${\overrightarrow{x}}$). For an observed past configuration ${\overleftarrow{x}} $, the future statistics are dictated by the conditional probability $p(\overrightarrow{X}=\overrightarrow{x}| \overleftarrow{X}=\overleftarrow{x}) $, which we abbreviate as $p(\overrightarrow{x}|\overleftarrow{x}) $.

By categorising all sets of past events with the same future statistics into equivalence classes (called causal states, which are encoded as memory states of the simulator), the optimal classical model (called the $\epsilon$-machine \cite{Crutchfield1989,Crutchfield1994}) only needs to store the class $\epsilon (\overleftarrow{x}) $ that $\overleftarrow{x} $ belongs to. That is, given only $\epsilon(\overleftarrow{x})$ the $\epsilon$-machine is able to make a statistically accurate inference of the process' conditional future. By observing the outcome of the stochastic process over a long time, one can infer the probability of each causal state and transition probabilities between them. For a stochastic process, the $ N $ causal states $\ S= \{ {S_i}\} _{i = 1}^N$ and their relevant transition probabilities are enough to realise the $ \epsilon $-machine model. The resulting $\epsilon$-machine requires~\cite{Book-Nielsen2010}

\begin{equation}
{C_{\mu}} = - \sum\limits_{i = 1}^N {{d_i}} \,\mathrm{log}\,{d_i},
\label{statisticalcomplexity}
\end{equation}

\noindent bits of information about the past, where $d_i$ is the probability that the past is in causal state $S_i$. No other predictive model can simulate the future while storing less information about the past. Thus $C_{\mu}$ has been termed the statistical complexity \cite{Zambella1988,Crutchfield1989,Crutchfield2012}, and is considered a fundamental measure of complexity that captures how resource-intensive it is to predict the future of a given process.

It has been theoretically proven that for many processes, including the one studied here, there exists a quantum $\epsilon$-machine with entropy $C_q$, such that $C_q < C_{\mu}$ \cite{Gu2012}.
Similar to its classical counterpart, this quantum model is defined by its causal states $\{|S_i\rangle\}$ and the corresponding transition probabilities. On average, the entropy of such a quantum $\epsilon$-machine is given by

\begin{equation}
{C_q} = - \mathrm{Tr}(\rho \log \rho),
\end{equation}

\noindent where $\rho = \sum\limits_i d_i {\left| {{S_i}} \right\rangle \left\langle {{S_i}} \right|} $.

\subsection*{Three-step simulation of a perturbed coin}
For the perturbed coin process, the optimal quantum causal states can be written as \cite{Gu2012}:

\begin{equation}
|S_0\rangle=\sqrt{l}\ |0\rangle +\sqrt{1-l}\ |1\rangle,
\end{equation}
\begin{equation}
|S_1\rangle=\sqrt{1-m}\ |0\rangle +\sqrt{m}\ |1\rangle.
\end{equation}

To give an example of the output state of our simulator, let us consider a perturbed coin defined by its parameters $l$ and $m$, which we denote as process $\Pi_1$. The output of the corresponding quantum $\epsilon$-machine after three time steps is given by the superposition
 \begin{equation}
\sum\limits_{x_n} {\sqrt {p\left( {{x_1},{x_2},{x_3}|{S_i},{\Pi_1}} \right)} } \left| {{x_1},{x_2},{x_3}} \right\rangle \left| {{S_{x_3}}} \right\rangle,
\label{eq:outputstate}
\end{equation}

\noindent where $ n=\{1, 2, 3\} $ and $p\left( {{x_1},{x_2},{x_3}|{S_i},{\Pi_1}} \right)$ is the probability to obtain $ x_1 $, $ x_2 $, and $ x_3 $ as the outcomes of three time steps of the process $\Pi_1$ when the input causal state is $ |S_i\rangle $. The value of $p$ can be evaluated theoretically from the transition probabilities between causal states (Fig.~\ref{fig:process}). The variables $ x_n \in \{0,1\} $ are the configurations of random variables $ x_1 $, $ x_2 $, and $ x_3 $, respectively. To sample from the future statistics of the perturbed coin process, we perform a simultaneous measurement of all the ancillary qubits after the three time steps. By also characterising the polarisation state of the photon in each case, we can tomographically reconstruct the output state associated with each time bin, and thus experimentally determine the statistical complexity of the simulation. To calculate the statistical complexity, ${C_q}$, for this process, we need to find the state $ \rho $:
\begin{equation}
\begin{split}
&\rho = {d_0}\sum\limits_{x_n} {{p( x_{1},x_{2},x_{3}|S_{0},{\Pi}_1)}\,\rho _\mathrm{{{pol}{|S_0}}}} \, \\
&+ {d_1}\sum\limits_{x_n} {{p( x_{1},x_{2},x_{3}|S_{1},{\Pi}_1)}\,\rho _\mathrm{{{pol}{|S_1}}}},
\end{split}
\label{eq:ensemble state}
\end{equation}

\noindent where

${d_0} = \frac{{\sum\limits_{{x_1},{x_2}} {{p({{x_1},{x_2},{x_3} = 0|{S_1},\Pi_1)}}} }}{{\sum\limits_{{x_1},{x_2}} {{p({{x_1},{x_2},{x_3} = 1|{S_0} ,\Pi_1)}}} + \sum\limits_{{x_1},{x_2}} {{p({{x_1},{x_2},{x_3} = 0|{S_1} ,\Pi_1)}}} }} $, $ {d_1}=1-{d_0} $, and $S _\mathrm{{{pol}{|S_i}}} $ is the tomographically reconstructed polarisation state at each arrival time, conditioned on the input memory state being encoded in $ |S_i\rangle$.

\subsection*{Verifying the unitarity of the processor via two-photon quantum interference}
 To verify that the operation is unitary, which guarantees the conservation of the entropy, we need to show that the superposition of different modes, both in time and polarisation, is coherent and that this coherence is maintained throughout the whole process. Using a pure state as the input and viewing the entire simulation as a black box, the output of the unitary operations inside the box should ideally be a pure state. In order to experimentally demonstrate this, we consider the case
 %
 %
 of simultaneously implementing two setups to model two identical processes, $ \Pi_1 = \Pi_2 $. It is possible to verify that two uncorrelated single photons are in identical pure states via two-photon interference---the Hong-Ou-Mandel (HOM) effect. The visibility of the interference, $v = \frac{ \mathrm{P}_{\max } - \mathrm{P}_{\min }}{ \mathrm{P}_{\max }} $, where $\mathrm{P}_{\max }$ ($\mathrm{P}_{\min }$) is the maximum (minimum) of two-photon coincidence detections measured when varying the delay between the two beams, can only be unity if the photons are in pure and identical states.

\subsection*{Comparison of future statistics}
The case of unequal processes also provides useful information. If $ \Pi_1\ne \Pi_2 $ and the output states are pure, the overlap of different future output statistics can be deduced by interfering the output photons.
For two photons in states $ |\psi\rangle $ and $ |\phi\rangle $ entering two input ports of a $ 50:50 $ beam splitter, the probability of finding a coincidence is $\frac{{1 - {{\left| {\left\langle {\phi } \mathrel{\left | {\vphantom {\phi \psi }}\right. \kern-\nulldelimiterspace}{\psi } \right\rangle } \right|}^2}}}{2}\ $, where $ \langle\phi|\psi\rangle$ is the overlap of the two states. Therefore, one can use the HOM interference visibility $v$ to estimate overlaps, by noting that $v = |\langle\phi|\psi\rangle|^2$. For our stochastic processes, the overlaps of the photonic output states are directly related to the overlaps of the future statistics produced by the two processes. For two different processes $ \Pi_1 $ and $ \Pi_2 $, let $ |S_i\rangle $ be a causal state of $ \Pi_1 $, and $ |T_j \rangle$ be a causal state of $ \Pi_2 $. Using Eq.~(\ref{eq:outputstate}), in general the overlap between the respective outputs of the quantum simulators for $\Pi_1$ and $\Pi_2$ will be
\begin{equation}\label{eq:6}
\sum_{x_n}\sqrt{p\left(x_{1},x_{2},x_{3}|S_{i},\Pi_{1}\right)p\left(x_{1},x_{2},x_{3}|T_{j},\Pi_{2}\right)}\langle S_{x_3}|T_{x_3}\rangle.
\end{equation}
 Since the perturbed coin process has Markov order one, and there is a one-to-one correspondence between the classical outcome and the causal state the machine transitions to, interfering the output states from a pair of quantum simulators for $\Pi_1$ and $\Pi_2$ as in Eq.~\eqref{eq:6}, actually results in an overlap
\begin{equation}
\sum_{x_n}\sqrt{p\left(x_{1},x_{2},x_{3} ,x_{4}|S_{i},\Pi_{1}\right)p\left(x_{1},x_{2},x_{3},x_{4}|T_{j},\Pi_{2}\right)}.
\end{equation}
That is, in this special case we are able to find the difference between conditional futures up to
one additional time step. Therefore, we can use our photonic quantum information processor for two tasks: (1) to simulate the future outcomes over three time steps of the classical stochastic process, and (2) to estimate the overlap of the future output statistics over four time steps.

\subsection*{Details of the experimental design}
The schematic in Fig.~\ref{fig:setup} shows how we implement the multi-step quantum-enhanced stochastic processor. Consider, for instance, the scenario where we want to sample the statistics. For one process (i.e. for one beam) a single photon is injected in the left-hand side of the circuit, from the source, with the state $\ket{0}$ $ (\ket{1})$ which is encoded as $\ket{H}=\ $horizontal ($\ket{V}=\ $vertical) polarisation. The first wave plate creates the desired initial causal state of our perturbed coin, either $\ket{S_0}_{\mathrm{pol}}=\sqrt{l}\ket{H}_{\mathrm{pol}}+\sqrt{1-l}\ket{V}_{\mathrm{pol}}$ or $\ket{S_1}_{\mathrm{pol}}=\sqrt{1-m}\ket{H}_{\mathrm{pol}}+\sqrt{m}\ket{V}_{\mathrm{pol}}$. The purpose of the first block is to transform a photon with a causal state encoded in polarisation into an appropriately weighted superposition of the classical outcomes of the first step encoded in the arrival time (denoted here as the delay degree of freedom, del), with the corresponding next causal state encoded in the polarisation:

\begin{equation}
\ket{S_0}_{\mathrm{pol}}\ket{0}_{\mathrm{del}}\rightarrow\sqrt{l}\ket{S_0}_{\mathrm{pol}}\ket{0}_{\mathrm{del}} +\sqrt{1-l}\ket{S_1}_{\mathrm{pol}}\ket{t_1}_{\mathrm{del}},
\end{equation}	
\begin{equation}
\ket{S_1}_{\mathrm{pol}}\ket{0}_{\mathrm{del}}\rightarrow\sqrt{1-m}\ket{S_0}_{\mathrm{pol}}\ket{0}_{\mathrm{del}} +\sqrt{m}\ket{S_1}_{\mathrm{pol}}\ket{t_1}_{\mathrm{del}}.
\end{equation}

This is achieved by temporarily using the photon path as an auxiliary degree of freedom:
A polarising beam splitter maps the polarisation degree of freedom onto the path, which is then copied onto the arrival time through the use of different path lengths ($ |H\rangle  \to  \mathrm{short\ path} $ and $ |V\rangle \to \mathrm{long\ path} $). By using a wave plate in each of the two paths, a path-dependent (and therefore, arrival-time-dependent) transformation of the polarisation into one of the two causal states is achieved: $ \ket{S_0}_{\mathrm{pol}}$ in the short path, and $ \ket{S_1}_{\mathrm{pol}}$ in the long path.

Next, the information on the path degree of freedom is erased, to avoid an exponential scaling of the number of paths (and optical elements in the experiment) with the number of time steps. To this end, the paths are recombined in a $50:50$ beam splitter, and subsequently post-selected for the photon exiting in the right output arm at the end of the first block (Fig.~\ref{fig:setup}). This means that we will lose half of our photons at the beam splitter, but in each run that we post-select, the evolution is unitary because the post-selection ensures that no photon is detected in the other output arm. By repeating the described block at each time step, we have three blocks to realise a three-step machine. The use of a sequence of interferometers has also been demonstrated in other experiments to study different topics in quantum information, such as  non-Markovian dynamics and sequential state discrimination~\cite{chiuri2012linear,nagali2012testing}.

To be able to attribute a different arrival time to each sequence of classical outcomes, we require a unique path length for every possible combination of short or long paths within the three blocks. The delays are implemented as $t_1=2$ ns at the first step, $t_2=4$ ns at the second, $t_3=8$ ns at the third step. The arrival times are discriminated by time-resolving single-photon detectors. The coincidence window for HOM interference is long enough to include the state which is spread out in a 14 ns time interval.

After the third step, we have the measurement stage at one output arm of the third BS and the circuit continues at the other, which is exploited for the second task of our work. In order to run our simulation and estimate the memory efficiency of this scheme compared to the optimal classical one, we measure the final arrival times (encoding the three ancillary qubits of the original scheme) and reconstruct the final polarisation state of the photon. This can be done simultaneously at the tomography stage, by also measuring the arrival times of the photons, allowing a full reconstruction of the polarisation state and arrival time.

The same apparatus can be exploited for the interference part of our experiment, the only difference being that now two single photons are injected in the setup. They both pass through the three blocks described above. When we want to verify the unitarity of our simulation, the elements in the blocks are the same for both the photons, so as to have $\Pi_1=\Pi_2$; on the other hand, they are different when we want to compare the future statistics of two different processes ($\Pi_1\ne \Pi_2$). After the output of the third block, the two photons interfere in a fibre BS and the number of coincidences is measured.

\subsection*{Details of the $ l $ and $ m $ parameters used in the experiment}

  The simulated process, for which the $ C_q $ results are depicted in Fig.~\ref{fig:result}a, is a perturbed coin with parameters $ l=0.4 $ and $ m $ ranging from $ 0.1 $ to $ 1.0 $ in increments of $ 0.1 $. Due to experimental imperfections the actual implemented values of $ l $ and $ m $ slightly deviate from the nominal ones ($ l=0.397 $ and $ m=\{0.101, 0.197, 0.297, 0.391, 0.490, 0.588, 0.685, 0.784, 0.882,\\ 0.994\} $). In Fig.~\ref{fig:result}c, the turquoise and magenta colours both show the case of two processes. For the turquoise graph, the fixed process is a stochastic process of a perturbed coin with input causal state $ |S_0\rangle $, $ l=1.0 $, and $ m=1.0 $. The varying stochastic processes are the ones with input causal state $ |S_0\rangle $, $ m=0.5 $, and nominal $ l=\{0.25, 0.50, 0.70, 0.85, 0.95, 1.0\}$ (the parameter $ l $ is used to change between different processes). For the magenta graph, the fixed stochastic process is a perturbed coin with input causal state $ |S_0\rangle $, $ l=0.5 $, and $ m=0.5 $. The varying ones are the stochastic processes with input causal state $ |S_0\rangle $, $ m=0.5 $, and nominal $ l=\{0.00, 0.10, 0.30, 0.50, 0.70, 0.90, 0.99\}$.


\section*{Acknowledgements}

We thank Raj B. Patel for helpful contributions. This research was funded, in part, by the Australian Research Council (project no.~DP160101911),
the Lee Kuan Yew Endowment Fund (Postdoctoral Fellowship), Singapore Ministry of Education Tier 1 grant RG190/17, Singapore National Research Foundation Fellowship NRF-NRFF2016-02, and NRF-ANR grant NRF2017-NRF-ANR004 VanQuTe, and the FQXi Large Grant: The role of quantum effects in simplifying adaptive agents. F.G. acknowledges support by the Australian Government Research Training Program (RTP) scholarship. We acknowledge the traditional owners of the land on which this work was undertaken at Griffith University, the Yuggera people.

\section*{Contributions}

 FG, NT and CDF designed the experimental setup; FG and NT performed the experiment and analysed the data. CDF, MG and JT conducted the theory of the project, as well as contributing to the data analysis. GJP played a significant role in the project conceptualisation, provided experimental assistance, and oversaw all aspects of the project. All authors contributed to writing the manuscript.

\end{document}